\documentclass[sigconf]{acmart}
\usepackage{tabularx}
\usepackage{booktabs}
\AtBeginDocument{%
  }

\setcopyright{none}
\setcopyright{none}
\renewcommand\footnotetextcopyrightpermission[1]{}

\begin{document}

\title{Reading Between the Lines: A Study of Thematic Bias in Book Recommender Systems}

\author{Nityaa Kalra}
\email{nityaak5@gmail.com}
\affiliation{%
  \institution{Radboud University}
  \city{Nijmegen}
  \country{The Netherlands}
}

\author{Savvina Daniil}
\email{s.daniil@cwi.nl}
\affiliation{%
 \institution{Centrum Wiskunde \& Informatica}
 \city{Amsterdam}
 \country{The Netherlands}
 }



\begin{abstract}
Recommender systems help users discover new content, but can also reinforce existing biases, leading to unfair exposure and reduced diversity. This paper introduces and investigates thematic bias in book recommendations, defined as a disproportionate favoring or neglect of certain book themes. We adopt a multi-stage bias evaluation framework using the Book-Crossing dataset to evaluate thematic bias in recommendations and its impact on different user groups. 

Our findings show that thematic bias originates from content imbalances and is amplified by user engagement patterns. By segmenting users based on their thematic preferences, we find that users with niche and long-tail interests receive less personalized recommendations, whereas users with diverse interests receive more consistent recommendations. These findings suggest that recommender systems should be carefully designed to accommodate a broader range of user interests. By contributing to the broader goal of responsible AI, this work also lays the groundwork for extending thematic bias analysis to other domains.

\end{abstract}





\maketitle
\pagestyle{plain}
\section{Introduction}

Recommender systems have become central to how we engage with the vast digital ecosystems of today. Whether it is finding the next movie to watch, discovering new music, or selecting a book to read, these systems not only help us navigate overwhelming choices, but actively shape our consumption habits \cite{Fleder2010}. In the context of books, recommendation algorithms influence the types of narratives and ideas to which users are exposed. This exposure can lead to \textit{psychological transportation}, a process known to influence individuals' attitudes, beliefs, and ultimately their interpretation of the world \cite{Green2000}. 

Socialogist Pierre Bourdieu's work \cite{Bourdieu1987} shows that cultural consumption is never neutral. Our preferences are shaped by underlying social structures, and in turn, they shape how we are perceived and judged by others. Further, Lawrence \cite{Lawrence2020} argues that reading preferences, if left unchecked, can become barriers to a fair and democratic society. From this perspective, recommending books is not merely a technical task; it is an ethical responsibility of librarians to actively promote diverse content rather than simply catering to individual desires. This argument becomes even more important in the context of book recommender systems, which mediate our reading choices without human intervention.

However, these systems are not neutral. They often favor popular titles \cite{Naghiaei2022} or authors from dominant demographic groups \cite{Daniil2022}, reinforcing existing cultural hierarchies. Although popularity and demographic bias have received increasing attention, this work turns to a less studied but equally important concern: \textbf{thematic bias}. Thematic bias refers to a content-level skew in which certain book themes are disproportionately favored, while others are overlooked. Unlike biases tied to user behavior or author identity, thematic bias emerges from the semantic and ideological framing of the content itself, reflecting what kinds of ideas and narratives are being amplified or marginalized. When algorithms amplify specific themes due to popularity dynamics or engagement metrics, they participate in shaping cultural discourse. This is visible in communities like BookTok and Bookstagram, where specific narratives often achieve viral dominance due to platform algorithms and influencer trends \cite{Balling2024, Ridzuan2023}. Therefore, studying thematic bias is not simply about improving the accuracy of personalization. Instead, it is about recognizing how these systems might reproduce or even intensify cultural inequalities.

When discussing thematic bias, it is important to clarify what we mean by themes. Traditionally, they refer to the underlying ideas or messages conveyed by a work \cite{Rimmon1995}. For example, a classic novel like \textit{Pride and Prejudice} addresses multiple themes such as love, social class, and individual growth. However, in this work, we define themes as clusters of semantically related terms derived from a large collection of book descriptions using topic modeling techniques. While this approach may not capture the full depth or complexity of a literary theme, it provides a practical and scalable means to represent thematic content across thousands of titles. It is worth noting that thematic bias does not arise in isolation. It is often shaped by structural patterns within both the data and the algorithms. For instance, Daniil et al. \cite{Daniil2022} show how demographic bias stems from popularity bias. Similarly, thematic bias may result from the dominance of certain user preferences or widely consumed content.

In this paper, we investigate whether thematic biases emerge in book recommender systems and how user preferences interact with these biases using a multi-stage bias evaluation framework. We adopt a segmentation strategy inspired by intersectionality, a concept originally developed to understand how overlapping social identities create unique experiences of privilege or disadvantage \cite{Bowleg2012, Cho2013, HillCollins1990}. By extending this idea to user modeling, we analyze how multiple dimensions of reading behavior, such as thematic diversity and propensity for popular items, combine to shape users’ recommendation experiences. To guide this investigation, we pose the following research questions:
\begin{enumerate}
    \item To what extent does thematic bias exist in the input data used by book recommender systems?
    \item How do different recommendation algorithms amplify or mitigate thematic biases in their outputs?
    \item How do thematic biases affect different user groups, particularly in relation to their thematic diversity and propensity for popular items?
\end{enumerate}

Our contributions include enriching the Book-Crossing dataset \cite{Ziegler2005} with thematic metadata via topic modeling, proposing a theme-based bias analysis pipeline, and applying an intersectional segmentation of users based on their reading behaviors. To the best of our knowledge, this is the first work to analyze thematic bias using topic modeling in book recommender systems with an intersectional user-group fairness lens. To support transparency and reproducibility of our work, all code, data, and analysis notebooks are publicly available on GitHub.\footnote{\url{https://github.com/nityaak5/Bias-Book-Recommendation}}

\section{Related Work}
\subsection{Bias in Recommender Systems}

While recommender systems enhance user experience by personalizing content, they are also susceptible to various biases that can lead to skewed, unfair, or homogenized recommendations. A widely documented phenomenon is \textbf{popularity bias}, where algorithms tend to favor already popular items. Abdollahpouri et al. \cite{Abdollahpouri2019} formalized this tendency within collaborative filtering (CF) algorithms. Further, Naghiaei et al. \cite{Naghiaei2022} demonstrated that these algorithms consistently under-served users with diverse or niche tastes, while performing well for bestseller-focused readers.

Recommendation outcomes are also shaped by deeper structural issues such as data-level and algorithmic biases. \textbf{Data-level bias }\cite{Chen2023} arises from the nature of the user-item interaction logs that recommender systems rely on. These logs reflect only a small and non-random subset of user preferences. As a result, the data tends to overrepresent items that were already popular or frequently shown, regardless of their actual relevance. The issues in the input data are further compounded by \textbf{algorithmic bias }\cite{Hajian2016}, particularly through feedback loops \cite{Mansoury2020} which can reinforce existing interaction patterns, leading to an echo chamber effect \cite{Noordeh2020}.

Several studies have documented the demographic implications of bias in recommender systems. For instance, collaborative filtering models have been shown to reflect, and in some cases amplify gender distribution found in user profiles, thereby reinforcing existing disparities in representation \cite{Ekstrand2021}. Other work has shown that such models may also favor books by U.S. authors over those from other regions \cite{Daniil2022}. While Ziegler et al. \cite{Ziegler2005} used the Book-Crossing dataset to examine topical diversity, their focus was on enhancing personalization rather than investigating thematic bias in recommendation systems. Initial steps to address these biases have been taken in adjacent domains such as music and movies, where researchers have proposed calibrated recommendation strategies to improve genre balance \cite{Steck2018}. This work builds on that gap by examining whether recommendation algorithms favor certain literary themes over others, and how such biases affect different user groups.

\subsection{Topic Modeling}

Understanding thematic content of text is essential in many Natural Language Processing (NLP) and Information Retrieval tasks. One of the most widely used approaches is topic modeling \cite{Vayansky2020}, which aims to uncover the thematic structure in a collection of documents. 

Traditional models such as Latent Dirichlet Allocation (LDA) \cite{Blei2003Latent} and Non-negative Matrix Factorization (NMF) \cite{Lee1999NMF} have been widely used to extract themes from textual data. These models represent each document as a mixture of topics, making them suitable for content-rich domains like books, where summaries and descriptions often convey multiple narrative dimensions. More recent advances in NLP have introduced embedding-based topic models that leverage pre-trained language models. For example, Top2Vec \cite{Angelov2020} and BERTopic \cite{Grootendorst2022} utilize word or sentence embeddings to represent documents in a semantic vector space, which are then clustered to identify coherent topics. 

In the context of recommender systems, topic models have been used to derive content features that improve recommendation quality, especially when user-item interaction data is sparse \cite{Wang2011}. However, the application of topic modeling in book recommendation research remains underdeveloped despite the richness of book metadata such as summaries, blurbs, and user-generated reviews. In this study, we utilized \textbf{BERTopic} \cite{Grootendorst2022} to extract fine-grained themes from book descriptions. These extracted themes formed the foundation of our analysis of \textbf{thematic bias} in recommendation outcomes.

\section{Research Design}

\subsection{Dataset Preparation}

For this study, we utilized the Book-Crossing dataset \cite{Ziegler2005} which originates from a four-week crawl of the \textit{BookCrossing}\footnote{\url{https://www.bookcrossing.com}} website. The raw dataset consists of over 1 million book ratings. However, it is highly sparse and noisy. For example, many books appear under multiple ISBNs due to edition changes, author names follow inconsistent formats, and most books have received very few ratings.

To reduce noise and improve robustness, we applied preprocessing steps similar to Naghiaei et al. \cite{Naghiaei2022}. We removed users with fewer than 5 or more than 200 ratings, books with fewer than 5 ratings, and all implicit ratings (i.e., rating = 0). We also merged duplicates using a combination of title and author fields.

To support theme-based bias analysis, we enriched books with textual descriptions and genre labels from external sources. While genre labels are available in some cases, we chose not to rely on them exclusively due to their broadness and inconsistency. At times, they fail to capture the detailed thematic information needed for effective bias analysis. For instance, many books are simply labeled \textit{Fiction} despite being clearly \textit{Psychological Thrillers} or \textit{Crime Thrillers}. Moreover, genre metadata is frequently missing for less popular titles, thereby limiting its coverage. To address these issues, we collected detailed book descriptions and later applied topic modeling to extract consistent thematic representations.

\begin{figure}[h]
    \centering
    \includegraphics[width=0.65\linewidth, height=10cm]{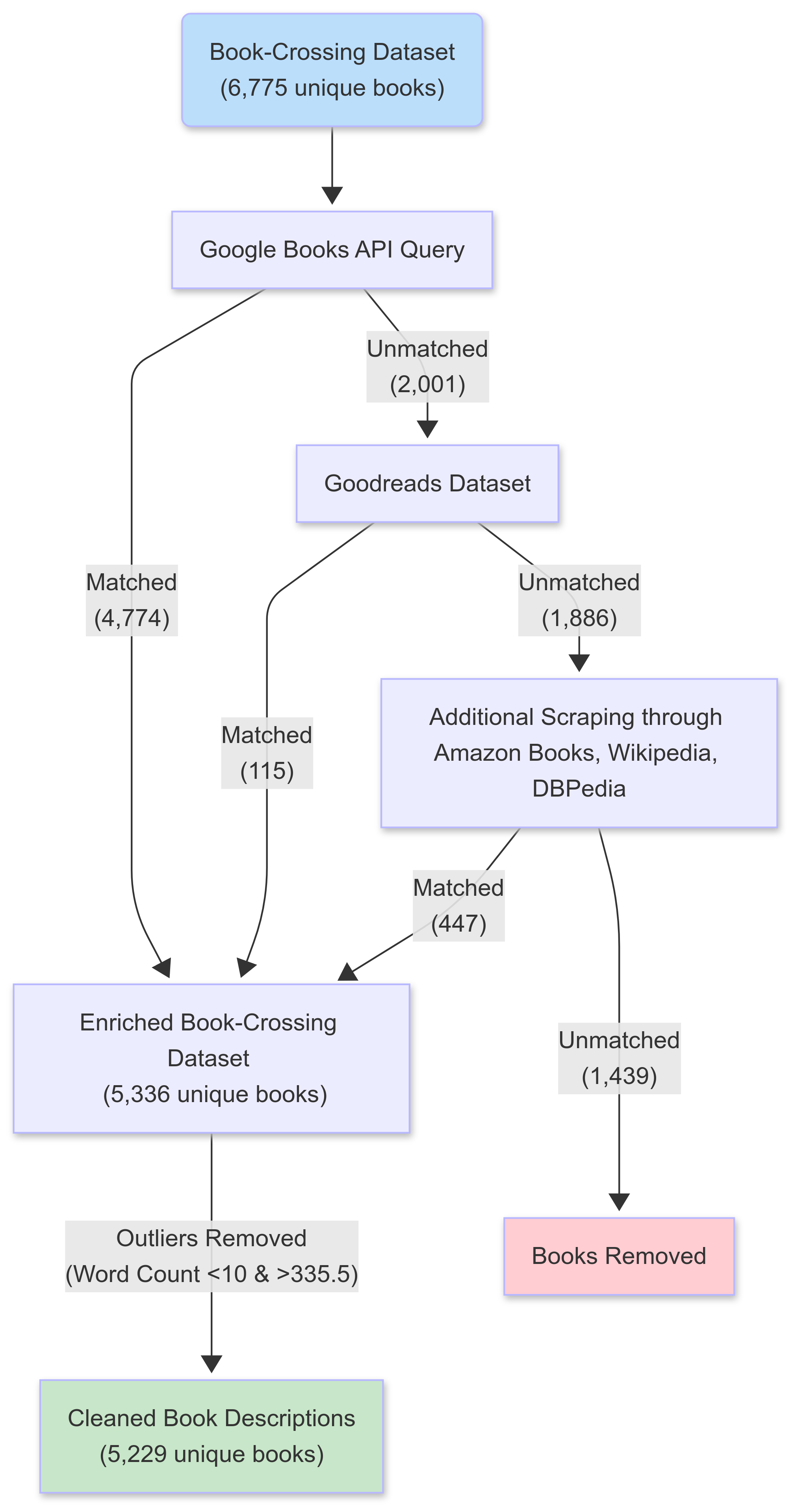}
    \caption{Metadata enrichment pipeline}
    \Description{Metadata enrichment process for the book-crossing dataset.}
    \label{fig:data-enr}
\end{figure}

 As shown in Figure \ref{fig:data-enr}, we first queried the Google Books API\footnote{\url{https://developers.google.com/books}} using ISBN or title-author combination, and retrieved metadata for approximately 2,000 books.
 Since Goodreads stopped issuing new API keys and discontinued public developer access in December 2020, we used a scraped version of the Goodreads dataset\footnote{\url{https://www.kaggle.com/datasets/jealousleopard/goodreadsbooks}} for the remaining books. We also performed additional scraping through Amazon Books\footnote{\url{https://www.amazon.com/books}}, Wikipedia, and DBpedia. Next, we cleaned book descriptions to remove outliers based on word count. Finally, books which could not be enriched via any of these methods were removed. Table \ref{tab:desc_stat} provides detailed statistics of the dataset, both before (without implicit ratings) and after pre-processing.

\begin{table}[ht]
\small
\centering
\caption{Dataset statistics before and after data enrichment}

\begin{tabular}{lcc}
\toprule
\textbf{Metric} & \textbf{Raw Dataset} & \textbf{Post-Enrichment} \\
\midrule
Users & 68,091 & 5,424 \\
Books & 149,836 & 5,229 (with Descriptions) \\
      &         & 4,984 (with Categories) \\
Interactions & 383,842 & 70,622 \\
Sparsity & 99.90\% & 99.75\% \\
Average Rating & 7.63 & 7.84 \\
Median Rating & 8.00 & 8.00 \\
Ratings/User & 5.64 & 13.02 \\
Ratings/Book & 2.56 & 13.51 \\
\bottomrule
\end{tabular}
\label{tab:desc_stat}
\end{table}

\subsection{Thematic Modeling}

To extract themes from the book descriptions, we used BERTopic \cite{Grootendorst2022}, a state-of-the-art topic modeling method that uses contextual embeddings from transformer models. 

We modeled 5,229 cleaned book descriptions with a minimum cluster size of 10 and cosine similarity threshold of 0.8 for topic merging. An initial run yielded 65 topics. After merging and pruning low-quality clusters, we retained 25 final topics and referred to them as \textit{themes}. Next, we manually reviewed them based on the top 10 keywords, sample book descriptions, and the label created using BERTopic’s inbuilt LLM functionality. Finally, labels were assigned to each of these themes to reflect the core narrative or subject matter in the book descriptions. 

To validate the quality of discovered themes, we combined internal and external checks. Internally, we evaluated them through manual inspection, stability across multiple runs, and topic coherence scores (average = 0.52). Externally, we compared them against genre metadata and reading trends from the early 2000s. Common themes such as \textit{Family Dynamics and Relationships}, \textit{Crime Thriller} and \textit{Detective Drama}, aligned well with bestseller patterns from \textit{The New York Times} \cite{nyt} and \textit{Publishers Weekly} \cite{pw}, while introspective themes such as \textit{Spiritual and Self-Help} matched post-9/11 literary commentary \cite{Self2014NovelDead}.
 This dual validation supported relevance of our themes for bias analysis.

\subsection{Bias Analysis Framework}
To investigate bias in book recommendation systems, we implemented a structured framework combining collaborative filtering (CF) algorithms with multi-stage bias evaluation.

We selected both memory-based and model-based approaches alongside two simple baselines. CF algorithms were chosen for their effectiveness in modeling user-item interactions and their known susceptibility to popularity bias \cite{Abdollahpouri2019, Naghiaei2022, Daniil2022}. Implementation was done using the Cornac library\footnote{\url{https://cornac.preferred.ai/}}, with an 80-10-10 train-validation-test split and hyperparameter tuning based on NDCG@10. The algorithm choices were aligned with prior work on bias in book recommender systems \cite{Naghiaei2022, Daniil2022} to ensure comparability of our results. Table \ref{tab:algorithms} lists all recommender algorithms used in our experiments.

\begin{table}[ht]
\centering
\caption{Recommender algorithms used in the study}
\small
\begin{tabular}{lll}
\hline
\textbf{Algorithm} & \textbf{Acronym} & \textbf{Approach} \\
\hline
Random & Random & Baseline \\
Most Popular & MostPop & Baseline \\
Matrix Factorization & MF & Matrix Factorization \\
Weighted Matrix Factorization & WMF & Matrix Factorization \\
Non-negative Matrix Factorization & NMF & Matrix Factorization \\
Probabilistic Matrix Factorization & PMF & Matrix Factorization\\
Hierarchical Poisson Factorization & HPF & Matrix Factorization \\
Bayesian Personalized Ranking & BPR & Pairwise Ranking \\
User-based K-Nearest Neighbor & User-kNN & Nearest Neighbors\\
Neural Matrix Factorization & NeuMF & Neural Networks\\
Variational Autoencoders& VAE-CF & Neural Networks \\
\hline
\end{tabular}
\label{tab:algorithms}
\end{table}

We computed the top-10 book recommendations for each user using the selected algorithms. These recommendations were then analyzed to assess three types of bias: \textbf{data bias}, \textbf{recommendation bias}, and \textbf{user group bias}.

To analyze user group bias, we first segmented users based on their propensity for popular books. Following the definition of popularity from \cite{Naghiaei2022}, we identified three groups: \textbf{Mainstream readers}, whose reading history consisted of more than 70\% popular books, \textbf{ Mixed readers}, with 30–70\% of their books from the popular set, and \textbf{Long-Tail readers}, with fewer than 30\% of their books classified as popular. Next, we segmented users by their historical thematic diversity and Gini ratio. Users below the 25th percentile in theme count and above the 75th percentile in Gini were labeled as \textbf{Specialist readers}, those above the 75th percentile in theme count and below the 25th percentile in Gini as \textbf{Generalist readers}, and the rest as \textbf{Moderate readers}. We then compared the top-10 recommendations of each group with their historical preferences.

\section{Results and Discussion}
\subsection{Bias in Data}

\begin{figure*}[ht]
        \centering
    \includegraphics[width=0.8\linewidth]{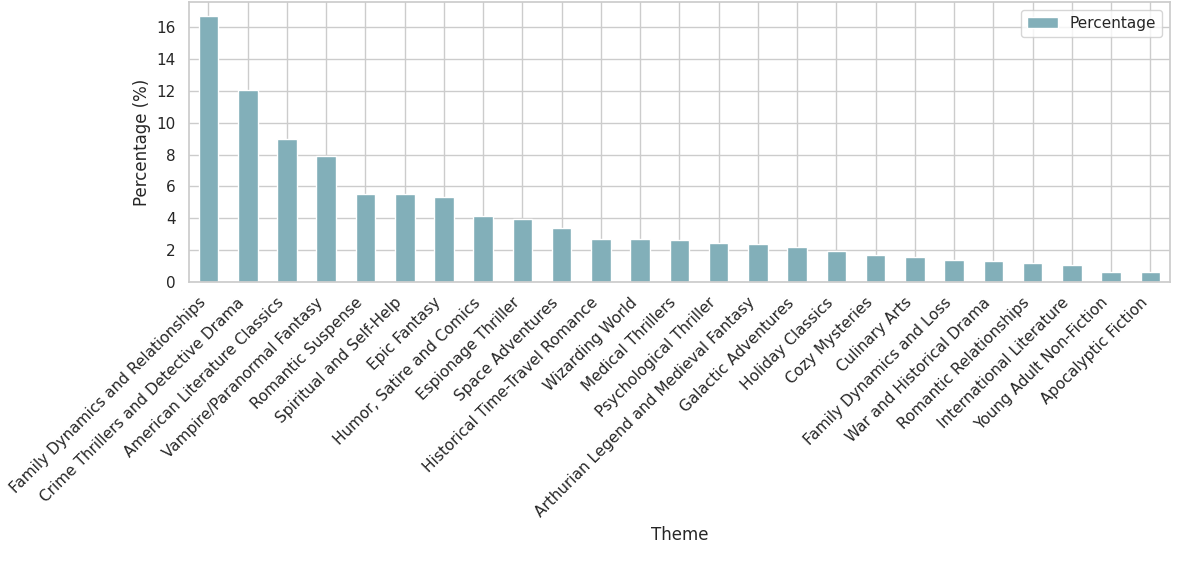}
        \caption{Distribution of themes among unique books in training dataset}
        \Description{A bar chart displaying percentage of each theme in a set of unique books.}
    \label{fig:themes}
\end{figure*}

We extracted \textbf{ 25 distinct themes} using BERTopic from book descriptions. The thematic distribution was highly imbalanced: about 20\% of themes accounted for over 52\% of unique books, revealing a clear imbalance in theme representation as seen in Figure \ref{fig:themes}.

\begin{figure}
    \centering
    \includegraphics[width=0.96\linewidth]{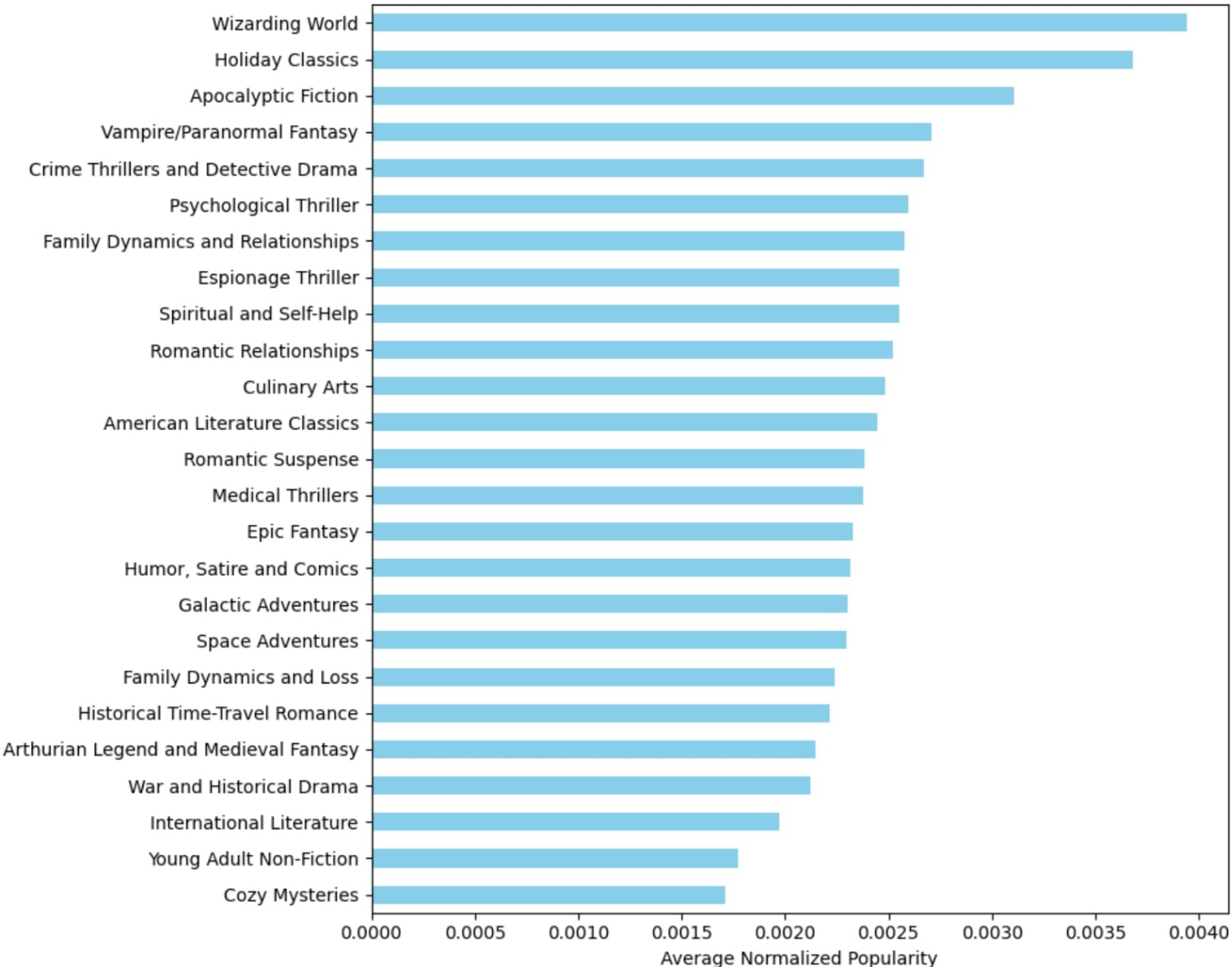}
    \caption{Average popularity ratio of themes}
    \Description{Average Popularity ratio of different themes calcluated by dividing popularity ratio of each book within a theme to number of books in a theme.}
    \label{fig:av_popularity}
\end{figure}

We defined a popular subset of books using the approach of Naghiaei et al. \cite{Naghiaei2022} and analyzed this subset to see which themes were overrepresented among popular books, both by comparing theme distribution and by computing the average popularity ratio of books within each theme. Several themes such as \textit{Crime Thrillers and Detective Drama}, \textit{Family Dynamics and Relationships}, and \textit{Vampire/ Paranormal Fantasy} showed clear overrepresentation among popular books. In contrast, themes such as \textit{Cozy Mysteries} and \textit{Space Adventures} were among the most underrepresented themes in the popular subset, despite of their moderate presence in the overall dataset. To determine if these differences were statistically meaningful, we conducted a
Chi-squared test of independence. The results showed statistically significant disparities (p < 0.05) in the distribution for 8 out of 25 themes. Furthermore, themes like \textit{Wizarding World} and \textit{Holiday Classics}, despite of their moderate representation in the training dataset, had exceptionally high average popularity ratios (Figure \ref{fig:av_popularity}), indicating engagement-driven amplification bias. This suggests that themes with fewer books can still exert disproportionate influence in recommendations if they attract high user engagement. 

\textbf{Therefore, in response to RQ1,} thematic bias is clearly present in the input dataset used by book recommender systems and it manifests as a multi-layered phenomenon. It stems from the initial imbalance in theme representation and is reinforced by user engagement patterns. Together, these biases create a compounding disadvantage for underrepresented themes and set the stage for potential downstream disparities in recommendation outputs.

\subsection{Bias in Recommendations}
To analyze recommendation bias, we first compared predictive accuracy of different algorithms using various metrics such as precision, recall, F1 score, and NDCG at k=10. WMF consistently outperformed all models across these metrics. VAECF and NeuMF also showed strong performance, particularly in balancing ranking and relevance. BPR and PF offered competitive performance, especially in Recall and F1, outperforming traditional factorization models. In contrast, user-KNN, MF and PMF performed poorly. Complete results across several other metrics are available in our Github repository.

Next, we analyzed how these models distributed themes in their \textbf{top-10} recommendations. Thematic recommendation percentages showed that \textit{Vampire/Paranormal Fantasy}, \textit{Family Dynamics and Relationships}, and \textit{American Literature Classics} dominated the output of most algorithms, mirroring their prevalence in the input data. In contrast, themes such as\textit{ Cozy Mysteries}, \textit{International Literature}, and \textit{War and Historical Drama} remained nearly absent in recommendations. We calculated the exposure ratio to quantify the difference between the prevalence of a theme in the dataset and the recommendations.

\begin{displaymath}
\text{\textit{Exposure Ratio}} = \frac{\% \text{ \textit{Theme in Recommendation}}}{\% \text{ \textit{Theme in Training Data}}}
\end{displaymath}
\\
\textit{Wizarding World} consistently exhibited a high Exposure Ratio among all algorithms. In fact, other themes that were overrepresented were those which had a high average popularity ratio in the first place (Figure \ref{fig:av_popularity}). This amplification highlights how themes with high engagement can be disproportionately promoted. We also evaluated item coverage across all algorithms. As expected, MostPop had the lowest coverage (0.19\%), while WMF stood out with the highest coverage (91.11\%), demonstrating its ability to explore a large portion of the item catalog. PMF showed moderate coverage (42.47\%), but other models like NeuMF, PF, BPR, and VAECF covered less than 15\% of the catalog, indicating a tendency to recommend from a small item pool and thereby limiting user exposure to diverse items.

\begin{figure*}
    \centering
    \includegraphics[width=0.69\linewidth, height=9.3cm]{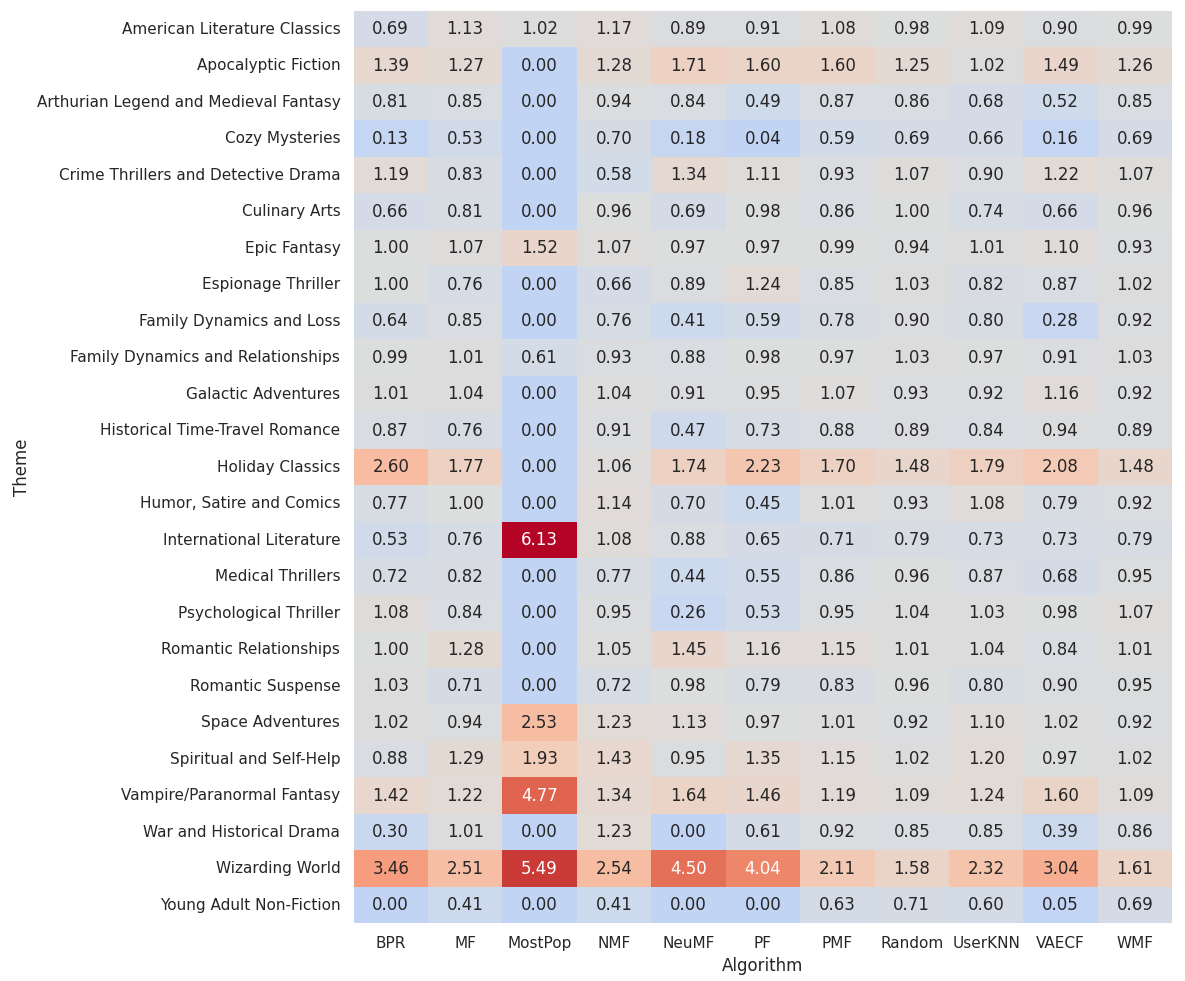}
    \caption{Exposure ratio across different themes and algorithms}
    \Description{Exposure ratio of themes calculated by dividing their percentages in recommendation to that in the original dataset}
    \label{fig:exposure}
\end{figure*}

\textbf{To answer RQ2}, we observe that most algorithms tend to amplify existing popularity biases. Models with high accuracy (WMF, VAECF) often overrepresent themes that were already popular in the input data. However, niche or underrepresented themes remain largely excluded, especially by models with low coverage and popularity-driven focus (MostPop, BPR). Even algorithms that achieve broader coverage (WMF) still contribute to thematic skewness by disproportionately favoring high-engagement content. Therefore, while predictive accuracy is not inherently tied to fairness, most algorithms do little to diversify exposure or correct imbalances in theme representation. 

\subsection{Bias Across User Groups}
\label{sec:bias-user-groups}

To analyze bias across different user groups, we first examined how recommendations varied based on user's propensity for popular items. A consistent trend was observed across all algorithms: \textbf{recommendations were heavily geared towards popular books, regardless of the user's historical preference}. This popularity bias was most pronounced among long-tail readers, whose preferences were overridden by mainstream recommendations. Interestingly, these users did not necessarily prefer niche themes, they simply engaged with less popular books within dominant themes, thereby challenging the assumption that long-tail users inherently seek niche content.

Next, we examined how recommendations differed for user groups defined by their historical thematic diversity. Specialist readers, with an average of 3.4 themes and a Gini index of 0.361, received more thematically diverse recommendations (6.8 themes, Gini reduced to 0.218), but this expansion often came with thematic displacement, as dominant themes in user profile shifted to \textit{Vampire/Paranormal Fantasy} or \textit{American Literature Classics}, depending on the algorithm. Generalist readers, with initially broad interests (10.2 themes, Gini 0.183), saw a reduction in diversity (to 7.2 themes) and were similarly steered towards dominant themes. Moderate readers showed mixed outcomes, some gained and others lost diversity. 
In summary, \textbf{despite starting from different levels of thematic diversity, both specialist and generalist readers were effectively normalized towards around 7 themes on average}, with a few themes being disproportionately represented. This algorithmic homogenization raises concerns about long-term user engagement and discovery. The results are shown in Table \ref{tab:gini}.

\begin{figure}[h]
    \centering
\includegraphics[width=0.8\linewidth, height=4.1cm]{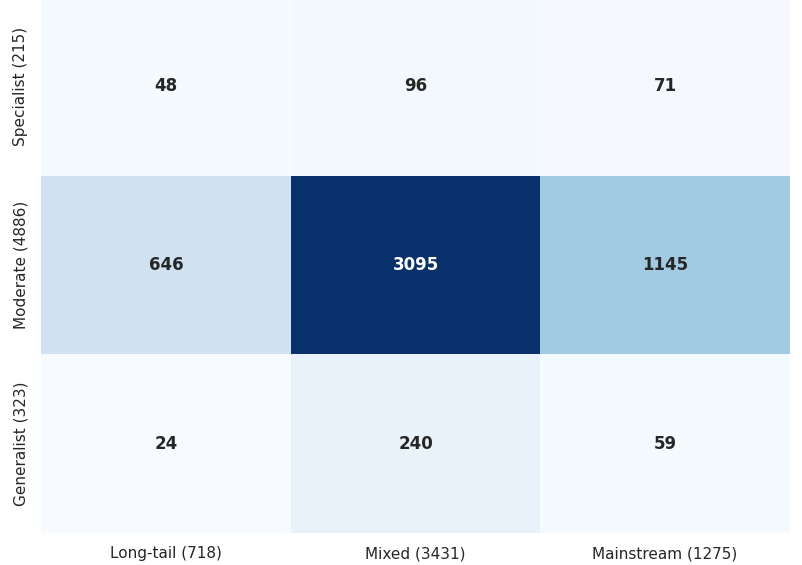}
    \caption{Intersectional subgroups }
    \Description{A heatmap showing division of user groups on 2 axis: thematic preferences and popularity orientation, resulting in 9 sub groups}
    \label{fig:heatmap}
\end{figure}

By intersecting users' thematic preferences and propensity for popular items, we identified nine subgroups (Figure~\ref{fig:heatmap}) and analyzed how their recommendations changed in terms of theme composition, popularity levels, and diversity. We observed that \textit{specialist readers with long-tail reading preferences} were one of the most affected user groups. For these users, recommended books frequently diverged from their historical preferences, both in terms of dominant themes and popularity levels. In contrast, \textit{moderate readers with mixed reading preferences} were least affected, with recommendation outputs remaining largely aligned with their past behavior. 

\begin{table}[h]
\small
\centering
\caption{Average themes and gini ratio across user groups}
\begin{tabular}{lcccc}
\toprule
\textbf{User Group} & \multicolumn{2}{c}{\textbf{Themes}} & \multicolumn{2}{c}{\textbf{Gini Ratio}} \\
 & \textbf{Data} & \textbf{Rec.} & \textbf{Data} & \textbf{Rec.} \\
\midrule
Specialist & 3.4 & 6.8 & 0.361 & 0.218 \\
Moderate & 6.8 & 7.1 & 0.226 & 0.204 \\
Generalist & 10.2 & 7.2 & 0.183 & 0.194 \\
\bottomrule
\end{tabular}
\label{tab:gini}
\end{table}

\textbf{Therefore, to answer RQ3,} thematic biases in recommendations impact users depending on their thematic diversity and propensity for popular items. We found a clear homogenizing effect where both specialist and generalist readers were steered towards a common range of themes. This homogenization was consistently biased towards a few dominant themes, displacing historical preferences. Our intersectional approach uncovered that users with both niche thematic preferences and long-tail consumption patterns experienced the highest degree of algorithmic skew towards popular items and dominant themes. This highlights that recommender systems can limit genuine discovery and reinforce echo chambers, especially for users with unique preferences.

\section{Conclusion and Future Work}

This work examined thematic bias in book recommender systems, focusing on how such bias arises from both data imbalance and algorithmic reinforcement. We enriched the Book-Crossing dataset with thematic metadata and proposed a framework to evaluate recommendation fairness through a theme-centric lens. Our analysis clearly demonstrated presence of thematic bias, driven by a two-fold phenomenon: an initial imbalance in theme representation within the dataset's unique book content, which is further reinforced by user engagement patterns. This sets a clear precedent for downstream recommendation behavior, as algorithms are more likely to learn from dominant themes.

Instead of evaluating bias along a single dimension, we jointly considered user's preferences for popular content and thematic breadth. In particular, users with niche and long-tail reading habits often received poorly aligned recommendations, as algorithms nudged them towards more mainstream content. In contrast, users whose preferences aligned with dominant patterns in the dataset received consistent recommendations. These findings make it clear that evaluating recommender systems through a universal lens is not enough. To truly support a wide range of users, we need to understand how different groups are impacted in different ways.

While our approach offers valuable insights, it comes with certain limitations. The topic modeling technique assigned only one theme to each book, even though many books naturally cover multiple themes. While this simplification made our analysis more manageable, it might have affected the granularity of bias detection. Future work could improve thematic labeling by using multi-label annotation or LLM enhanced topic modeling techniques such as the one by Kapoor et al. \cite{Kapoor2024}. Additionally, we focused solely on collaborative filtering algorithms without extensive fine-tuning and evaluated only the top-10 recommendations. Future studies could explore content-based or hybrid methods and assess performance across broader recommendation ranges. Finally, due to lack of directly comparable literature addressing thematic bias in book recommender systems, our findings largely stood alone without direct benchmarks for comparison. Looking ahead, applying the framework to other content domains such as movies, music, or podcasts could reveal how thematic bias differs across industries and whether common patterns emerge. Exploring the long-term effects of repeated exposure to biased recommendations on user behavior and engagement would highlight the broader consequences of bias and reinforce the urgency of developing effective bias-mitigation strategies.
 
\bibliographystyle{ACM-Reference-Format}
\bibliography{sample-base}

\end{document}